\documentclass[doublecol]{epl2}
\title{Control spiral wave dynamics using feedback signals from line detectors }
\shorttitle{Control spiral wave dynamics using feedback signals from line detectors} %Insert here a short version of the title if it exceeds 70 characters

\author{Guoyong Yuan\inst{1,2}\footnote{g-y-y1975@sohu.com}, Aiguo Xu\inst{3}, Guangrui Wang\inst{3} \and Shigang Chen\inst{3}}% \and S. Author\inst{1} \and T. Author\inst{2}}
\shortauthor{G.Y.Yuan \etal}

\institute{
  \inst{1} Department of Physics,Hebei Normal University - Shijiazhuang,050016, China\\
  \inst{2} Hebei Advanced Thin Films Laboratory -
  Shijiazhuang,050016, China\\
\inst{3} Institute of Applied Physics and Computational Mathematics-
P. O. Box 8009, Beijing 100088,China}

\abstract{ We numerically study trajectories of spiral-wave-cores in
excitable systems  modulated proportionally to the integral of the
activity on the straight line, several or dozens of equi-spaced
measuring points on the straight line, the double-line and the
contour-line. We show the single-line feedback results in the drift
of core center along a straight line being parallel to the detector.
An interesting finding is that the drift location in $y$ is a
piecewise linear-increasing function of both the feedback line
location and time delay. Similar trajectory occurs when replacing
the feedback line with several or dozens of equi-spaced measuring
points on the straight line. This allows  to move the spiral core to
the desired location along a chosen direction by measuring several
or dozens of points. Under the double-line feedback, the shape of
the tip trajectory representing the competition between the first
and second feedback lines is determined by the distance of two
lines. Various drift attractors in spiral wave controlled by
square-shaped contour-line feedback are also investigated. A brief
explanation is presented. }
\begin{document}

\maketitle

\section{Introduction}
Spiral waves are typical examples of spatiotemporal patterns in
macroscopic systems driven far from thermodynamic equilibrium. They
exist extensively in excitable and self-oscillating media[1]. For
example,  cardiac muscle[2], platinum with oxidation of CO [3],
liquid crystal subjected to electric or magnetic field[4],the slime
mould dictyostelium discoideum[5] and reacting chemical systems[6].
The dynamics of spiral wave has attracted considerable interest,
which is attributed not only to the characteristics of its
nonlinearity and being far from equilibrium, but also to its
extensive destructions/applications. For example, spiral waves and
spatiotemporal chaos from repetitious breakup of spiral waves in
cardiac muscle may be the leading mechanism of tachycardia and
ventricular fibrillation[7]. Spiral waves in the brain are believed
to be associated with epilepsy[8]. Therefore, how to control or
eliminate spiral waves and spatiotemporal chaos is an interesting
research topic[9-14].

Recently, much attention has been paid to the dynamics of spiral
waves subjected to a feedback signal. The motions of spiral core
have been studied under a number of different feedbacks. These
feedbacks can be grouped into two classes, local feedback and
non-local feedback. Local feedback is also referred to as
one-channel feedback. It is performed according to the state at a
single measuring point of the excitable system. Non-local feedback
is performed according to states at more than one points, normally
the mean state over a region. Under local feedback control the
trajectory of spiral core is attracted to a series of limiting
cycles centered on the measuring point[15-21]. If the motion on the
innermost limit cycle is epicycloidal,  then this limiting cycle is
referred to as the entrainment attractor. Limiting cycles being
further from the measuring point are referred to as resonance
attractors. The radius of the attractor is dependent on the time
delay. The most straightforward generalization of one-channel
feedback is to monitor the states of system at two points and put in
signals from both to determine the strength of the feedback
signal[22-23].

It has been observed that two-channel feedback destroys the regular
dynamics seen in one-channel feedback if the measuring points are
sufficiently separated, and that several complex regimes appear when
varying the distance between the two measuring point. In the other
types of non-local feedback, the modulation is proportional to the
integral of the activity in 2D excitable domains of different
shapes[24-32]. The behaviors of attractors have been examined for
circular, square, elliptical, triangle, pentagon and rhombus
domains. When the domain size is significantly smaller than the
wavelength of spiral wave, the spiral core trajectories are similar
to those for local feedback. However, for larger domains, the
attractors depend significantly on the size and shape of the domain.

In this letter we numerically study  the behaviors of feedback
control from several kinds of line detectors to compensate the lack
of one-dimensional(1D) case between the zero-dimensional point and
the 2D domains mentioned above. The measuring domain is one of the
following cases, (i) a straight line, (ii)several or dozens of
equi-spaced measuring points on the straight line , (iii)a
double-line, (iv)a contour-line.

%The organization of the paper is as follows:
%In Sect. II we describe the mathematical model of the excitable media under a feedback signal derived from a measuring line.
%In Sect. III we investigate
%the effects of a feedback from the measuring along a straight line.
%Behaviors of the attractors under a feedback signal from the
%measuring along a double-line and a contour line of square domain
%are studied in Sect.IV and Sect.V, respectively. Section VI presents
%a brief explanation and conclusion.

\section{The mathematical model}
In our study the FizHugh-Nagumo(FHN) model[33-34] is used to
describe the excitable media. The FHN model is a set of two-variable
``reaction-diffusion" equations. This model is generic for excitable
systems and can be applied to a variety of systems. It can reproduce
many qualitative characteristics of electrical impulses along nerve
and cardiac fibers, such as the existence of an excitation
threshold, relative and absolute refractory periods, and the
generation of pulse trains under the action of external currents.
The FHN model with a line feedback reads
\begin{eqnarray}
\frac{\partial u}{\partial
t}&=&\frac{1}{\varepsilon}[-v-u(u-a)(u-1)]+\nabla^{2}u+\frac{I}{\varepsilon}\\
\frac{\partial v}{\partial t}&=&-\gamma v+\beta u-\delta
\end{eqnarray}
where the variables $u(x,y,t)$ and $v(x,y,t)$ are the activator and
inhibitor, respectively; $\varepsilon$ is the time scale and named
excitability parameter; $a$ represents the threshold for excitation;
$\gamma$, $\beta$ and $\delta$ are parameters controlling the rest
state and dynamics. Here, $a=0.03$, $\gamma=1.0$,
$\delta=0.0$,$\beta=2.0$ and $\varepsilon=0.004$. To perform the
line feedback, the modulation signal $I(t)$ is computed as
\begin{equation}
I(t)=k_{fb}(B(t-\tau)-B_{0})
\end{equation}
where
\begin{equation}
B(t)=\frac{1}{L}\int_{L}u(x',y',t)ds'
\end{equation}
Thus, the feedback signal is proportional to the integral value $B$
of the first variable over the measuring line of length $L$ taken
with a time delay $\tau$. The parameter $k_{fb}$ is the feedback
gain, and $B_{0}$ is the average value of $B(t)$ over one revolution
of a spiral wave without feedback. In numerical simulations the
system (1)-(2) is integrated by split operator method with the time
step $\Delta t=0.005$ t.u., the space step $h=0.1 $ s.u. and a
$200\times200$ array. No-flux boundary conditions are used.

\section{Resonant drift of spiral wave under a straight-line feedback}

\begin{figure}
\includegraphics[width=8.5cm,height=7cm]{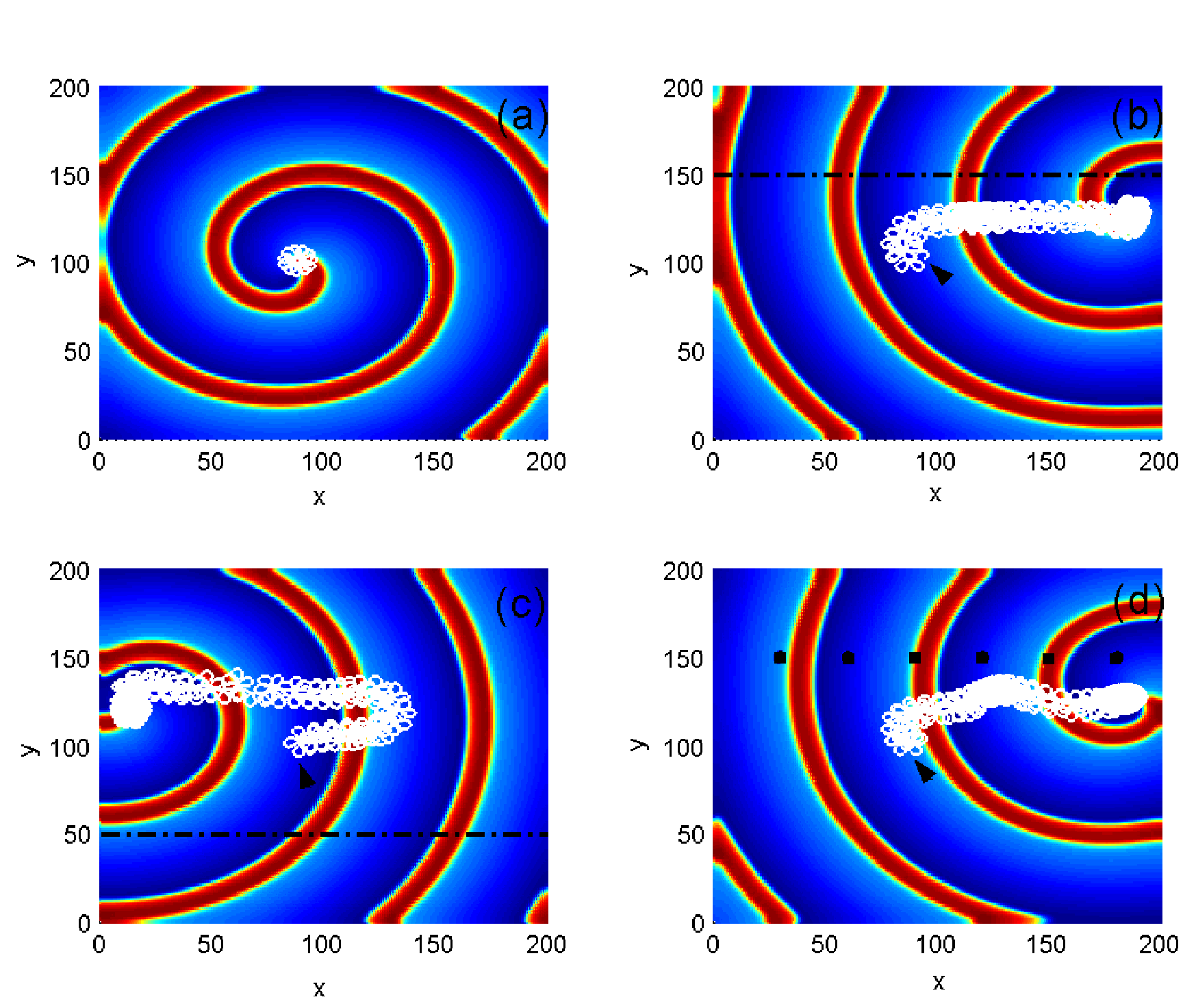}
\caption{\label{fig:wide} (Color online) Trajectories of a spiral
wave tip subjected to the feedback control shown in Eqs.(3) and (4)
for different locations of the line. Fig. (a) is for the initial
spiral wave and its tip before switching on the feedback control.
The feedback lines are indicated by back dash-dotted line in (b) and
(c). $y_{fb}$=150 in (b) and $y_{fb}$=50 in (c). Fig. (d) is for the
trajectory of a spiral wave tip subjected to the feedback derived
from six equi-spaced measuring points with $y_{fb}$=150. The spiral
images are shown for the end of the trajectory in (b),(c) and (d).
Here $k_{fb}=0.05$, $\tau=10$.}
\end{figure}

\begin{figure}
%\centerline{\epsfig{file= Fig2.ps, bbllx=128 pt,bblly=0 pt,bburx=655
%pt,bbury=522 pt, width=0.4\textwidth,clip=}}
\includegraphics[width=8.5cm,height=7cm]{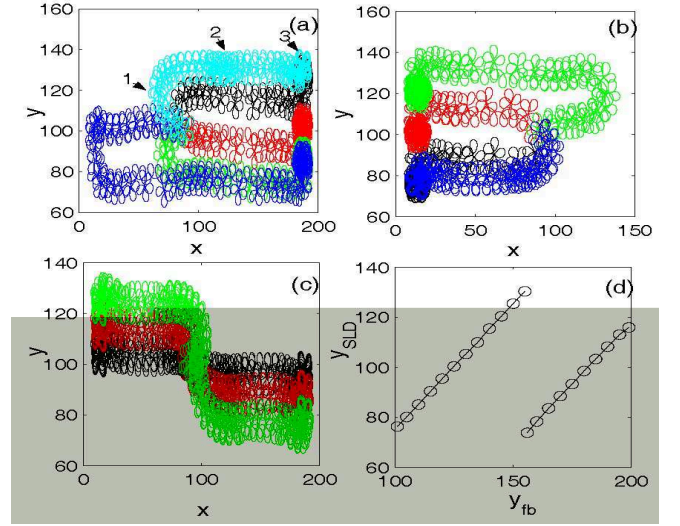}
\caption{(Color online)Trajectories of spiral wave tip under
feedback control for different locations of the feedback line. In
Fig. (a) the values of $y_{fb}$ corresponding to the black, red,
green, blue and cyan lines are 199, 175, 158, 156, and 155,
respectively. In Fig. (b) the values of $y_{fb}$ corresponding to
the black, red, green, blue,  lines are 5, 30, 50, 55, respectively.
In Fig. (c) the values of $y_{fb}$ corresponding to the left black,
left red, left green, right black, right red, right green are 80,
90, 100, 120, 112, 101, respectively. Fig.(d) shows the location of
the SLD $y_{SLD}$ vs. that of the feedback line $y_{fb}$. Here the
measuring line is parallel to the $x$-axis and has a length $L=200$.
 $k_{fb}$=0.05 and $\tau=10$.}
\end{figure}

\begin{figure}
\includegraphics[width=8.5cm,height=4cm]{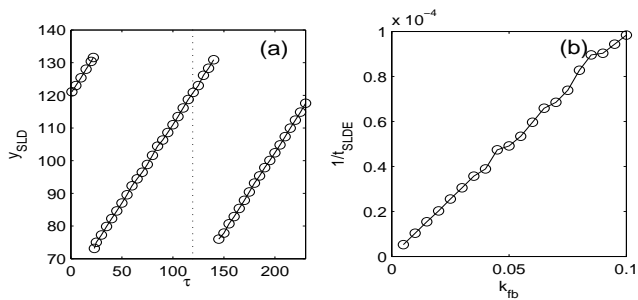}
\caption{\label{fig:wide} (a) Location $y_{SLD}$ of the SLD
trajectory vs. time delay $\tau$, where $k_{fb}=0.05$. (b)
$1/t_{SLDE}$ vs. the feedback gain $k_{fb}$,  where $t_{SLDE}$ is
the time of the SLD ending over and inversely proportional to the
drift velocity, $\tau=10$.}
\end{figure}

\begin{figure}
\includegraphics[width=8.5cm,height=4cm]{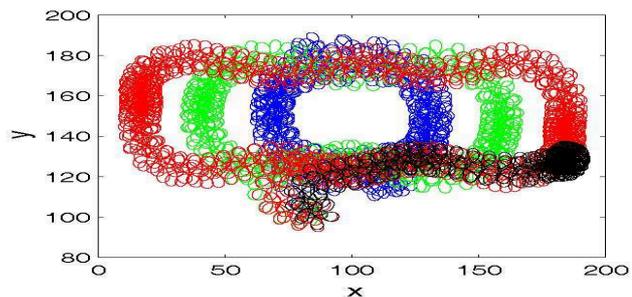}
\caption{\label{fig:wide} (Color online) Trajectories of a spiral
wave tip subjected to the feedback control for different lengths,
$L$, of the line with the fixed location, $y_{fb}=150$. $L=$40, 100,
160 and 180 for the blue, green, red and black lines. Here the
center of the measuring lines is fixed at $(100,150)$. $\tau=10$,
$k_{fb}=0.05$. }
\end{figure}

A single spiral can be induced from the equation system (1)-(2)
without the feedback term by truncating a traveling pulse. The
variables $u$ and $v$ are set initially to zero uniformly in the
medium. To create a spiral wave,  a super-threshold value $u=0.8$ is
given along a line near the boundary of the excitable medium to
induce a propagating wave. When the propagating wave approaches the
center of the excitable medium, one-half of the planar wave is
erased by resetting $u=v=0$. Subsequently, the open end of the
planar wave curls into a spiral wave with its core located near the
center of the excitable medium. The generated spiral wave rotates
with a rotation period $T_{0}\approx0.58$ .t.u. (about 116 time
steps) and a wavelength $\lambda\approx$ 5.8 s.u. (about 58 space
grids)(shown in Fig.1a). In the following parts of the paper, the
units of time and space are time step and spatial grid,
respectively.

Switching on the feedback control induces a drift of spiral wave
core. In the system described by the equation (1) and (2), under a
single point feedback, the spiral wave core has a drift along a
stable circular orbits centered at the measuring point. The initial
position of the detector with respect to the spiral tip, the time
delay and the feedback gain make effects on the nature of the
attractor, which is consistent with the feedback-controlled results
from the open gel reactor and the Oregonator model[17-18]. Now we
focus mainly on the resonant drift of spiral wave core subjected to
a line feedback. Considering the long measuring line which is
parallel to the $x$-axis and whose ends are on left and right
boundaries, respectively, we find that the drift trajectory induced
by the feedback signal is divided into three parts(shown in Fig.1
and Fig.2). In the part of trajectory labeled by 1 in Fig.2(a), the
spiral wave core drift first outwards from the domain center to a
certain location. In part 2 the trajectory starts a resonant drift
from a certain location to the boundary along a straight line
parallel to the measuring line, and finally stay near the boundary
with the form of the complex meandering attractor in part 3. The
final position of spiral core is close to either the left boundary
or the right one, which is determined by the location $y_{fb}$ of
the measuring line. The straight line drift(SLD),which is defined as
the part of the drift parallel to the measuring line and labeled by
2 in Fig.2(a), towards the right boundary occurs when $y_{fb}>100$
and the SLD is towards the left boundary when $y_{fb}\leq100$. In
Figs.1(b) and 1(c) the movement directions of SLD mediated by the
feedback derived from the line integral located at $y_{fb}=150$ and
$y_{fb}=50$ are compared. From Fig.2 it is found that, when the
absolute distance $|y_{fb}-100|$ increases, the part of trajectory
labeled by 1 rotates around the the initial spiral core in
counter-clockwise direction. Fig.2 (d) shows also that, when
$y_{fb}>100$, the location $y_{SLD}$ of the SLD part of the
trajectory,which is computed by the arithmetic mean of the maximum
and minimum y-values for the SLD part, is a piecewise
linear-increasing function of $y_{fb}$. On each linear part the
distance $y_{fb}-y_{SLD}$ has a constant value. The value for the
left linear part is about 25, while it is about 82 for the right
one.

The feedback is applied with a time delays $\tau$. As can be seen in
Fig.3(a), the value of  $\tau$ affects the location $y_{SLD}$ of
SLD. The function $y_{SLD}(\tau)$ is $T_{0}$-periodic and piecewise
linearly increasing. Further to increase or decrease the feedback
gain $k_{fb}$ does not result in a change in the drift trajectory of
spiral core. However, it affects the drift velocity of the core.
Fig.3(b) shows the relation between
 $1/t_{SLDE}$ and $k_{fb}$.
  $t_{SLDE}$ is the time interval from the start
of the drift to the formation of the final meandering attractor. It
is inversely proportional to the drift velocity.  $1/t_{SLDE}$ is
nearly linearly rising with the increase of the $k_{fb}$. The
effects of the length of the feedback line on a spiral wave are
shown in Fig.4. When the line length $L$ is significantly smaller
than the spiral wavelength $\lambda$ (about 58 space grids), the
spiral core finally follows a circular path with the center of the
short measuring line. The trajectories are similar to those for
local feedbacks. When the length increases, the circular path is
stretched along horizontal direction. If two ends of the line is
close to left and right boundaries, respectively, a transient SLD
and a final complex meandering attractor occurs.

Replacing the feedback line with several or dozens of equi-spaced
measuring points on the line, i.e., replacing the integral (4) with
the sum $\frac{1}{N}\sum_{n=1}^{n=N}u(x'_{n},y'_{n},t)$, we observe
the same drift trajectory as the line, where $N$ is the total number
of the measuring points. Fig.1(d) shows the trajectory of the spiral
tip with a six-points feedback, where the points lie collectively on
the line locating at $y_{fb}=150$ and have an equi-spaced
distribution from the left boundary to the right one. In this case,
the spiral tip follows the trajectory including a SLD toward left
boundary and a complex meandering attractor, which is nearly the
same as the one shown in Fig.1(b). It has become widely accepted
that the most dangerous cardiac arrhythmias are due to spiral waves
or reentrant waves. Therefore,it is very important to control spiral
waves in those systems. The above information tells that the
feedback signal derived from several or dozens of measuring points
can move the spiral core to the desired location along a chosen
direction.

\section{Resonant drift of spiral wave under a double-line feedback}
\begin{figure}
\caption{\label{fig:wide} (Color online) Trajectories(black) of a
spiral wave tip under the double-line feedback for the different
locations of the second measuring line. $y_{fb2}=105$ in (a).
$y_{fb2}=140$ in (b). $y_{fb2}=160$ in (c). $y_{fb2}=165$ in (d).
$y_{fb2}=175$ in (e). $y_{fb2}=185$ in (f). Here, the position of
the first measuring line is fixed at $y_{fb1}=150$, and
$y_{fb2}>100$. Red (green) curves show the trajectories when only
the first(second) feedback line exists. $L_{1}=200$, $L_{2}=200$,
$\tau=10$ and $k_{fb}=0.05$. }
\end{figure}

\begin{figure}
\caption{\label{fig:wide} (Color online) Similar to Fig.5, but here
$y_{fb2}\leq100$. $y_{fb2}=5$ in (a). $y_{fb2}=15$ in (b).
$y_{fb2}=30$ in (c). $y_{fb2}=60$ in (d). $y_{fb2}=65$ in (e).
$y_{fb2}=90$ in (f).}
\end{figure}

\begin{figure}
\includegraphics[width=8.5cm,height=6cm]{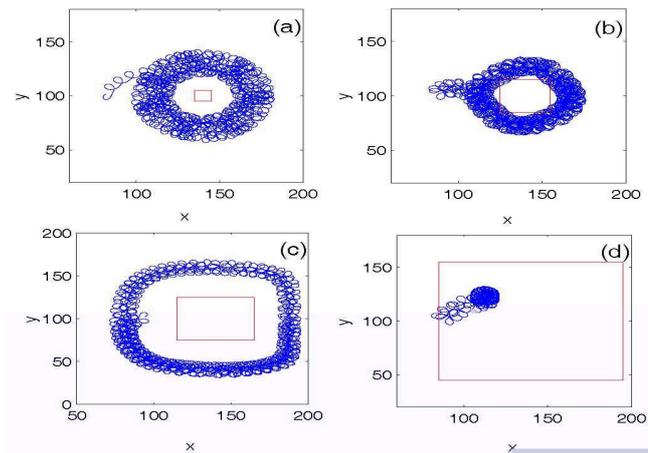}
\caption{\label{fig:wide}(Color online) Trajectories of a spiral
wave tip under various sizes of the square contour. Side length
$d=10$ in (a). $d=30$ in (b). $d=50$ in (c). $d=110$ in (d). The
contours are indicated by squares. $\tau=10$, $k_{fb}=0.05$.}
\end{figure}

The effects of the double-line feedback on a spiral wave are
examined by systematically varying the distance between two parallel
measuring lines, as shown in Fig.5 and Fig.6. Suppose that the lines
are parallel to $x-$axis, whose locations are described by $y_{fb1}$
and $y_{fb2}$, respectively. In the simulations we fix $y_{fb1}=150$
and change $y_{fb2}$. Figure 5 shows six typical examples of spiral
tip trajectories found for different values of $y_{fb2}$ under the
condition $y_{fb2}>100$. When the position $y_{fb2}$ approaches to
$100$, the first part of the trajectory with double-line feedback is
similar to the corresponding one of $y_{fb}=y_{fb1}$ in the
single-line feedback. Because the drift velocity with a single-line
feedback for $y_{fb}$ near $100$ is much
 smaller than that for $y_{fb}$ far away from $100$. After the
 transient period the trajectory forms directly a meandering attractor and doesn't
undergo a SLD, as shown in Fig.5a. When $y_{fb2}>130$, the SLD part
appears in the trajectory. A consecutive increase of $y_{fb2}$
results in a rotation in counter-clockwise direction for the first
of the drift and a periodic variety of the SLD location. Because the
same change occurs under a single-line feedback. A special case is
observed for $y_{fb2}=175$ due to the symmetry of singe-feedback
trajectories for $y_{fb}=150$ and for $y_{fb}=175$, where the
trajectory forms a meandering attractor near the initial core after
undergoing a transitional circular drift(see Fig.5e). $y_{fb2}=175$
is also a transition position in comparing the SLD location under
the double-line feedback with that under the single-line feedback
$y_{fb}=y_{fb2}$.

Now we examine the drift behavior under a double-line feedback with
$y_{fb2}\leq100$. When the second feedback line is close to the
bottom boundary, the first part of the trajectory with the
double-line feedback approaches that for the case with only the
second line. See Figs.6(a) and 6(b). Because the drift velocity is
large when the feedback line is close to the upper or lower boundary
in the single-line feedback control. On the other hand, for a large
$y_{fb2}$ being close to $100$, there is a first part approaching
that of the single-line feedback $y_{fb}=150$, see Fig.6(f). The
reason is due to the small drift velocity when $y_{fb}$ is close to
$100$. For moderate $y_{fb2}$, the drift includes the first part,
whose location lies between the corresponding part of the
single-line feedback $y_{fb}=y_{fb1}$ and that of $y_{fb}=y_{fb2}$,
the long or short SLD, and the final meandering, see Figs.6(c)-(e).
The SLD towards the left or right depends on the competition between
the trajectories with the single-line feedback $y_{fb}=y_{fb1}$ and
with $y_{fb}=y_{fb2}$.

\section{Resonant drift of spiral wave under a contour-line feedback}

The role of domain shape has been studied in relation to
reaction-diffusion systems with global feedback. Square-, circular-,
triangular- and elliptical-shaped domains are commonly used in
experiments and computations as they are simplest two-dimensional
confined geometries. In this section the influence of contour-line
of the feedback domains on the evolution of spiral waves is
investigated in detail.

Figure 7 shows the drift trajectories that correspond to four
square-shape contour lines with different side-lengths and a fixed
center (140,100). When the side-length $d$ is significantly smaller
than the spiral wavelength $\lambda$, the spiral leaves the initial
center by drifting toward the left until it turns to follow a
circular path, as shown in Fig.7(a). This motion resembles those
observed in the reported results applying a point feedback. In this
case, when $d$ increases, the radius of the resonance attractor
becomes larger slowly. Around $d=30$, the motion is changed, where
the spiral core first drifts away from the initial core, then, it
approaches a stable square trajectory, which is rotated by about
$45^{\circ}$, see Fig.7(b). As a result of the significant increase
of $d$, the spiral core is attracted toward a stable
squared-trajectory with a large side-length and an orientation
coinciding with the feedback contour, as shown in Fig.7(c). In the
range where this kind of attractors create, the size of the square
trajectory is slowly reduced with the increasing of $d$. An
interesting cross-shaped trajectory is generated when increasing the
contour-line size to $d=90$. This trajectory can be considered as a
combination of four small pieces of square trajectories linked
together. With a further increase of $d$ to make the left side of
the contour-line approach the initial core, a final complex
meandering trajectory locates inside the contour, see Fig7.(d).

Keep the center of  contour at $(90,100)$ close to that of the
initial spiral $(88,100)$, we observe the trajectories of spiral
wave tip under various side-lengths $d$ of the square contour. As
$d$ increases, the trajectory is similar to the contours with the
center at $(140,100)$. The difference is that, after the stable
square trajectory with orientational angle $45^{\circ}$ with respect
to the feedback contour, the complex meandering attractor locating
at the initial core appears.

% Spiral wave dynamics under feedback control derived from a square
% domain has been discussed in the reference [27]. By comparing the
% square-shaped contour and the corresponding domain with the contour
% as the boundary, we consider that the effects of  the domain
% feedback on spiral wave dynamics are mainly attributed to its
% contour-line feedback.

\section{Discussion and conclusion}
The analysis of spiral wave dynamics under feedback control via a
line may be performed via the drift velocity mediated by the
corresponding feedback. Here the dynamics can be understood via the
results of one-point feedback.

When $k_{fb}>0$, the drift induced by  feedback  can be expressed
as[32]
\begin{equation}
\gamma(x,y)=\varphi+\omega\tau+\phi(x,y)
\end{equation}
where $(x,y)$ is the Cartesian coordinates of the core center of
spiral unperturbed spiral wave, $\omega$ is the fundamental
frequency in the Fourier expansion of the feedback signal. When a
one-point feedback is applied, the phase of feedback signal follows
\begin{equation}
\phi(x,y)=\pi+arctan(\frac{y}{x})+\frac{2\pi}{\lambda}\sqrt{x^{2}+y^{2}}
\end{equation}
 for any points $(x,y)$ except the measuring point $(0,0)$.
 The drift velocity field is obtained from the phase $\phi(x,y)$ and
Eq.(5). The domain center is a stable fixed point, since within its
vicinity drift vectors are oriented toward the origin. A radial
displacement of the core center is accompanied by a counterclockwise
rotation of the vectors. At a certain displacement the vectors are
orientated perpendicularly to the radial direction. It turns out
that the circle with  certain radius is a limiting cycle of the
drift velocity field. The limiting cycle represents the so-called
resonance attractor of spiral wave.

Let us assume that the drift of spiral core is mainly affected by
the points which are in the feedback line and are the nearest from
the core center. The spiral first leaves the position where its core
was initially placed by drifting until it reaches the circular
attractor of these points. After traveling a short distance, the
neighboring points start to become the new nearest points. The
spiral core turns to follow the attractor of the neighboring points,
which may be obtained by translating the former attractor. This
process is continued until the spiral core reaches the vicinity of
boundary, and the SLD part being parallel to the feedback line
forms. When the tip approaches the boundary, the feedback signal is
periodic because  the feedback line crosses with the spiral wave, as
shown in Fig.1. So the complex meandering attractor is generated in
the domain close to the boundary. Replacing the feedback line with
several or dozens of equi-spaced measuring points on the line, the
same drift trajectories are observed. It may be explained by the
similarities between the attractor of one-point feedback and that of
the short-line feedback with the length significantly smaller than
the spiral wavelength.

Based on the above explanations, the distance $|y_{SLD}-y_{yb}|$
corresponds to the radius $R$ of the resonance attractor under
one-point feedback, which can be obtained from Ref.[32] to be
\begin{equation}
\frac{R}{\lambda}=m-0.25sign(k_{fb})-\frac{\varphi}{2\pi}-\frac{\tau}{T_{\infty}}
\end{equation}
where $m$ is an integer. By transforming $y_{SLD}$ to
$|y_{SLD-}y_{yb}|$, it is easy to find that Figs.2(d) and 3(a) agree
qualitatively with the theoretical predictions given by Eq.(7). For
each value of $\tau$ there are several possible stable attractors
corresponding  to different values of the integer $m$, which results
in the piecewise  behavior of drift lines versus feedback lines and
time delay.

In this letter,we discussed also the dynamics of spiral wave under
the double-line feedback and square-shaped contour one. The
trajectory of spiral tip with the double-line feedback may be
understood via the competition between two corresponding single-line
feedbacks. A variety of attractors can be obtained from a
contour-line feedback.

\acknowledgments This work is supported by the National Natural
Science Foundation of China [under Grant Nos.10647127 and 10775018],
Science Foundations of Hebei Education Department[under Grant
No.2009135], Hebei Nature Science Foundation Project[under Grant No.
A2006000128] and Science Foundation Of Hebei Normal University.


\begin{thebibliography}{99}

\bibitem{b.a}
  \Name{Cross M C. \and Hohenberg P C.}
  \REVIEW{Rev. Mod. Phys.}{65}{1993}{852}.

\bibitem{b.a}
  \Name{Courtemanche M}
  \REVIEW{Chaos.}{6}{1996}{579}.

\bibitem{b.a}
  \Name{Nettesheim S.,Oertzen A V.,Rotermund H H. \and Ertl G.}
  \REVIEW{J. Chem. Phys.}{98}{1993}{9977}.

\bibitem{b.a}
  \Name{Frisch T.,Rica S.,Coullet P. \and Gilli J M.}
  \REVIEW{Phys. Rev.Lett.}{72}{1994}{1471}.

\bibitem{b.a}
  \Name{Frisch C A.,Panfilov A V.,Hogeweg P.,Siegert F \and Weijer C J.}
  \REVIEW{J.theor.Biol.}{181}{1996}{203}.

\bibitem{b.a}
  \Name{Winfree A T. \and Strogatz S H.}
  \REVIEW{Physca.D}{8}{1983}{35}.

\bibitem{6a}\Name{Gray R A.,Jalife J.,Panfilov A V.,Baxter W T.,Cabo C.,Davidenko J M.\and Pertsov A.M.}
  \REVIEW{Science}{270}{1995}{1222}.

\bibitem{6a}\Name{Huang X.,Troy W C.,Yang Q.,Ma H.,Schiff S J.,\and Wu J Y.}
  \REVIEW{J.Neuroscience}{24}{2004}{9897}.

\bibitem{b.a}
  \Name{Kim M., Bertram M., Pollmann M., Oertzen A., Mikhailov A S., Rotermund H H. \and Ertl G.}
  \REVIEW{Science}{292}{2001}{1357}.

\bibitem{b.a}
  \Name{Sakaguchi H. \and Fujimoto T.}
  \REVIEW{Phys. Rev. E}{67}{2003}{067202}.

\bibitem{b.a}
  \Name{Wang P Y. and Xie P.}
  \REVIEW{Phys. Rev. E}{61}{2000}{5120}.

\bibitem{b.a}
  \Name{Zhang H., Cao Z., Wu N J., Ying H P. \and Hu G.}
  \REVIEW{Phys. Rev. Lett.}{94}{2005}{188301}.

\bibitem{b.a}
  \Name{Yuan G Y.,Wang G R. \and Chen S G.}
  \REVIEW{Europhys.Lett.}{72}{2005}{908}.

\bibitem{b.a}
  \Name{Yuan G Y., Chen S G. and Yang S P.}
  \REVIEW{Eur.Phys.J.B}{58}{2007}{331}.

\bibitem{b.a}
  \Name{Steinbock O.,Zykov V S. \and M$\ddot{u}$ller S C.}
  \REVIEW{Nature}{366}{1993}{322}.

\bibitem{b.a}
  \Name{Grill S., Zykov V S. \and M$\ddot{u}$ller S C.}
  \REVIEW{J. Phys. Chem.}{100}{1996}{19082}.

\bibitem{b.a}
  \Name{Braune M. \and Engel H.}
  \REVIEW{Phys. Rev. E.}{62}{2000}{5986}.

\bibitem{b.a}
  \Name{Grill S., Zykov V S. \and M$\ddot{u}$ller S C.}
  \REVIEW{Phys. Rev. Lett .}{75}{1995}{3368}.

\bibitem{b.a}
  \Name{Karma A \and Zykov V S.}
  \REVIEW{Phys. Rev. Lett .}{83}{1999}{2453}.

\bibitem{b.a}
  \Name{Zykov V S., Kheowan O U., Rangsiman O. \and M$\ddot{u}$ller S C.}
  \REVIEW{Phys. Rev. E.}{65}{2002}{026206}.

\bibitem{b.a}
  \Name{Kheowan O U., Zykov V S., Rangsiman O. \and M$\ddot{u}$ller S C.}
  \REVIEW{Phys. Rev. Lett .}{86}{2001}{2170}.

\bibitem{b.a}
  \Name{Goldschmidt D M., Zykov V S. \and M$\ddot{u}$ller S C.}
  \REVIEW{Phys. Rev. Lett .}{80}{1998}{5220}.

\bibitem{b.a}
  \Name{Zykov V S., Brandtst$\ddot{a}$dter H.,Bordiougov G. \and Engel H.}
  \REVIEW{Phys. Rev. E.}{72}{2005}{065201R}.

\bibitem{b.a}
  \Name{Zykov V S., Bordiougov G., Brandtst$\ddot{a}$dter H., Gerdes I. \and Engel H.}
  \REVIEW{Phys. Rev. Lett.}{92}{2004}{018304}.

\bibitem{b.a}
  \Name{Zykov V S. \and Engel H.}
  \REVIEW{Phys. Rev. E.}{70}{2004}{016201}.

\bibitem{b.a}
  \Name{Zykov V S., Mikhailov A S. \and M$\ddot{u}$ller S C.}
  \REVIEW{Phys. Rev. Lett.}{78}{1997}{3398}.

\bibitem{b.a}
  \Name{Kheowan O U., Zykov V S. \and  M$\ddot{u}$ller S C.}
  \REVIEW{Phys. Chem. Chem. Phys.}{4}{2002}{1334}.

\bibitem{b.a}
  \Name{Kheowan O U., Kantrasiri S., Wilairat P., Storb U. \and M$\ddot{u}$ller S C.}
  \REVIEW{Phys. Rev. E.}{70}{2004}{046221}.

\bibitem{b.a}
  \Name{Kheowan O U., Kantrasiri S.,Uthaisar ., G$\acute{a}$sp$\acute{a}$r V. \and M$\ddot{u}$ller S C.}
  \REVIEW{Chem.Phys.Lett.}{389}{2004}{140}.

\bibitem{b.a}
  \Name{Kheowan O U., Chan C K., Zykov V S., Rangsiman O. \and M$\ddot{u}$ller S C.}
  \REVIEW{Phys. Rev. E.}{64}{2001}{035201}.

\bibitem{b.a}
  \Name{Naknaimueang S., Allen M A. \and M$\ddot{u}$ller S C.}
  \REVIEW{Phys. Rev. E.}{74}{2006}{066209}.

\bibitem{b.a}
  \Name{Zykov V S. \and Engel H.}
  \REVIEW{Physica D.}{199}{2004}{243}.

\bibitem{b.a}
  \Name{FitzHugh R., et al}
  \REVIEW{Biophys. J.}{1}{1996}{445}.

\bibitem{b.a}
  \Name{Courtemanche M., Skaggs W. \and Winfree A T.}
  \REVIEW{Physica D.}{41}{1990}{173}.

%% \bibitem{b.b}
%%  \Name{Author F. \and Author S.}
%%  \Book{Some Book of Interest}
%%  \Editor{A. Editor}
%%  \Vol{9}
%%  \Publ{Publishing house, City}
%%  \Year{1939}
%%  \Page{666}.

%%\bibitem{b.c}
%%  \Editor{Editor A.}
%%  \Book{Some Book of Interest}
%%  \Vol{9}
%%  \Publ{Publishing house, City}
%%  \Year{1939}
%%  \Section{A}.
\end{thebibliography}
\end{document}